\documentclass[%
 reprint,
nofootinbib,
 amsmath,amssymb,
 aps,
]{revtex4-2}

\usepackage{graphicx}
\usepackage{dcolumn}
\usepackage{bm}
\usepackage[dvipsnames]{xcolor}

\usepackage{float}

\begin{document}

\preprint{APS/123-QED}

\title[Neutron star injections]{Probing Formation of Double Neutron Star Binaries around 1mHz with LISA
}


\author{Lucy O. McNeill}
\thanks{email: mcneill@tap.scphys.kyoto-u.ac.jp
}

\author{Naoki Seto}%
\affiliation{Department of Physics, Kyoto University, Kyoto 606-8502, Japan
}%

\date{\today}

\begin{abstract}
We propose a novel method to examine whether  Galactic  double neutron star binaries are formed in the LISA band.
  In our method, we assign  an effective time  fraction $\tau$ to each double neutron star binary detected by LISA.  This fraction is given as a function of the observed orbital period and eccentricity and should be uniformly distributed  in the absence of in-band binary formation. Applying  statistical techniques such as the Kolmogorov-Smirnov test to the actual list of $\tau$, we can inspect the signature of the in-band binary formation.  We discuss the prospects of this method, paying {close} attention to the available sample number of Galactic double neutron star binaries around 1mHz.

\end{abstract}

\maketitle


\section{Introduction}
Double neutron star binaries (hereafter DNSBs) are bountiful astrophysical targets.  They have been detected as radio pulsars \cite{Hulse1975,Manchester1977}, and  the  current sample has orbital periods $P$ from 1.9 hours \cite{Stovall2018} to six weeks \cite{Swiggum2015}. This sample contains around 20 DNSBs, and searches might be incomplete at the faint end of the luminosity function  (also limited by  the beaming fraction).
Shorter period ($P<$ 1 hour) DNSBs ought to exist in the Galaxy as well. However, due to Doppler smearing and shorter merger timescales, their detection would be more difficult than the longer period ones \cite{Bagchi2013}.

 The Laser Interferometer Space Antenna  (LISA) is planned in the 2030's \citep{Amaro2017} and is sensitive to gravitational waves (GWs) around 0.1-100mHz.
It will detect all Galactic DNSBs in the frequency range $f\gtrsim$1.5mHz  (corresponding to the orbital period  $P=2/f\lesssim 20$\,min),  unlike the longer-period radio sample. Observationally motivated estimates \citep{Kyutoku2019} suggest that at least dozens of DNSBs exist in the Galaxy  at $f \gtrsim$1.5mHz. Numerical galactic modelling \cite{Lau2020,Wagg2022} predicts that LISA will detect altogether  from a { few}  to upto hundreds of DNSBs.

{DNSB formation depends on the complex interplay between many astrophysical processes. In the general picture, the binary must survive two supernova explosions. Preceding these supernovae, various mechanisms have been theorized with respect to mass loss/exchange in the binary \cite{Paczynski1976,Podsiadlowski1992,Vink2001,Dewi2005} after hydrogen burning. These processes all play a role in} determining the separation at formation, if the binary survives.

However, the related efficiencies and rates in populations are {not well established} (see \cite{Mandel2022} for a review).
In particular, it's unclear whether there is a critical minimum orbital period (or maximum  orbital frequency) for isolated DNSB formation.

Dynamical encounters in star clusters is an alternative pathway for short period DNSB formation, {though} their contribution to the LISA sample is estimated  to be small {\cite{Phinney1991,Kremer2018}}. Also, for the dynamical scenario it will be difficult to  solidly estimate the distribution function for the orbital periods of the generated DNSBs.

In this paper, we are  interested in the possibility of DNSB formation specifically  at $f\gtrsim1$mHz. We hereafter call this channel  as  ``in-band" (mHz) DNSB formation, or simply ``injections".  Considering the aforementioned theoretical uncertainties, it  will be  fruitful to observationally examine the in-band formation in a model independent manner.

We thus develop a statistical method to examine the in-band formation with LISA  (see also \cite{Andrews2020} for formation between the LISA band and the lower frequencies already probed by the radio sample).  Recently, several studies have proposed to statistically deal with multiple LISA sources in the Galaxy.
Among others, a large number ($\sim10^4$) of  white dwarf  binaries  (WDBs) will be a powerful data set for various astronomical analyses (e.g., probing the Galactic structure \citep{Korol2019, Wilhelm2021}).  In this context, one of the authors suggested to measure the fluxes of  the Galactic WDBs in frequency space \cite{Seto2022}.  He pointed out that the measurement will enable us to follow  the collective evolution of the WDBs, resulting in mergers or stable mass  transfers.

 One might imagine that we can get some information about the in-band formation by similarly measuring the DNSB flux with LISA  at various  frequencies.  Unfortunately, LISA will detect much fewer DNSBs than WDBs, and {the small number statistics} will severely limit the  flux approach for DNSBs.
On the other hand,  unlike WDBs, DNSBs  can be well regarded as point particle systems in the LISA band, and their long-term orbital evolution from GW emission can be predicted quite accurately \cite{Hulse1975}.
Considering these pros and cons of DNSBs, we newly introduce the effective time  fraction $\tau$, corresponding to    the fraction of total time that each binary has spent in the mHz band.
Without in-band formation, the fraction $\tau$ should be uniformly distributed. {By analysing the observed list of $\tau$,  we can examine potential in-band formation through {its} deviation from a uniform distribution}.

The basic assumption of our study is that the mHz DNSB population is in the ``steady state" \cite{Kyutoku2016}.
This is a reasonable assumption, since a DNSB passes through  the mHz band in the timescale of Myr but the Galactic DNSB merger rate will change in the Hubble timescale {\cite{Nelemans2004,Lamberts2019}}.

This paper is organized as  follows. In section  \ref{sec:binaries}, we roughly estimate how many Galactic DNSBs are  likely to be  detected by LISA.  In section \ref{sec:lisaband} we  study the long-term orbital evolution of DNSBs and   define the effective time fraction $\tau$, as a function of the GW frequency  and the orbital  eccentricity.  Then, in section \ref{sec:injections},   using statistical tools such as the Kolmogorov-Smirnov test, we discuss how well we can probe in-band formation with LISA.  In section \ref{sec:caveats}, we mention potential extensions of this study. We summarize our findings in section~\ref{sec:summary}.
\section{Galactic double neutron star binaries}
\label{sec:binaries}
\subsection{Expected number in the LISA band }
\label{sec:MW-number}
We first estimate the merger rate $R_\mathrm{MW}$ of DNSBs in our Galaxy (Milky Way).  As a basic observational  input, we use the comoving merger  rate ${\mathcal{R}}=660 ^{+1040}_{-530}{\rm Gpc^{-3}yr^{-1}}$ from a recent report by the LVK collaboration \cite{BNS2022} (Multi source model).

To relate the two rates $\cal R$  and  $R_{\rm MW}$, we apply the traditional argument {based} on the effective number density of Milky Way equivalent galaxies {\cite{Phinney1991,Kalogera2001}}, and put $R_{\rm MW}=L_{B,\mathrm{MW}}{ \cal R}/\mathcal{L}_B$.   Here $\mathcal{L}_B$  is the B-band luminosity per comoving volume and $L_{B,\mathrm{MW}}$ is the B-band luminosity  of our Galaxy.
Using their typical values, we obtain
\begin{align}
R_{\rm MW} =& 6.0 \times 10^{-5}  \mathrm{yr}^{-1}
 \left( \frac{\mathcal{R}}{660\mathrm{Gpc}^{-3}\mathrm{yr}^{-1}} \right) \nonumber\\ & \times \left( \frac{L_{B,\mathrm{MW}}}{9\times 10^{9}L_\odot} \right) \left( \frac{\mathcal{L}_B}{10^{17} L_\odot\mathrm{Gpc}^{-3}}\right)^{-1}.
 \label{eq:RMW}
\end{align}
   We should notice that the Galactic merger rate  $R_\mathrm{MW}$ still has large uncertainties (at  least a factor  of  three).

   Next, we roughly
    estimate the total number of Galactic DNSBs in the LISA band. In this paper, we use  the notation $f$ specifically for the second harmonic frequency  (given by $f=2  f_\mathrm{orb}$ with  the orbital frequency $ f_\mathrm{orb}$).  Due to radiation reaction, the GW frequency $f$ evolves as \cite{Peters1964}

\begin{equation}
    \frac{\mathrm{d}f}{\mathrm{d}t} =  \frac{96 \pi^{8/3}G^{5/3}f^{11/3}\mathcal{M}^{5/3}}{5c^5 (1-e^2)^{7/2}}\left( 1+\frac{73}{24}e^2+\frac{37}{96}e^4\right)
    \label{eq:fdot}
\end{equation}
   with the orbital eccentricity $e$ and the chirp mass  $\cal  M$ of the binary \citep{Peters1964}.  At the same time, eccentricity evolves according to
   \begin{equation}
       \frac{\mathrm{d}e}{\mathrm{d}t} =  -\frac{304e \pi^{8/3}G^{5/3}f^{8/3}\mathcal{M}^{5/3}}{15c^5 (1-e^2)^{5/2}}\left( 1+\frac{121}{304}e^2\right).
    \label{eq:edot}
   \end{equation}
   These equations are a good approximation for DNSBs in the LISA band {since} the relativistic corrections are small (except for $1-e\ll 1$).

   Simply assuming (i) the steady state condition for the Galactic DNSB population at $f\gtrsim 1$mHz, and (ii) no binary formation there, we have the frequency distribution
\begin{equation}
    \frac{\mathrm{d}N}{\mathrm{d}f}= R_{\rm MW} \left( \frac{\mathrm{d}f}{\mathrm{d}t} \right)^{-1} \propto R_{\rm MW} {f^{-11/3}}{\mathcal{M}^{-5/3}}.
    \label{eq:dNdf}
\end{equation}
Here we ignored the eccentricity dependence of $\dot{f}$. After the frequency integral, we obtain the cumulative number
   \begin{equation}
\begin{split}
    N(>f) &= 30\left( \frac{\mathcal{M}}{1.2M_\odot} \right)^{-5/3} \left( \frac{f}{1.5\mathrm{mHz}}\right)^{-8/3}\\
    & \times \left( \frac{R_{\rm MW}}{6.0 \times 10^{-5}\mathrm{yr}^{-1}}\right).
    \end{split}
    \label{eq:Ng}
\end{equation}
 Note that the chirp mass distribution of known DNSBs is  narrow and centred around $\mathcal{M} = 1.2M_\odot$ \citep{Tauris2017}.

In fact, we will later relax the assumption (ii).  But the above result will serve as a rough guide, except for extreme model settings.

 \subsection{Gravitational wave observations}
 \label{sec:GW-observations}
 \subsubsection{Identification of binary neutron stars}
Next, we consider a DNSB at distance $D$, with chirp mass $\mathcal{M}$, eccentricity $e$ and gravitational wave frequency $f$. When the binary is approximated as monochromatic ($f=$ constant), the dimensionless gravitational wave strain amplitude from the binary in the second orbital harmonic is
\begin{equation}
    h_2 = \frac{8 G^{5/3}\mathcal{M}^{5/3}\pi^{2/3}f^{2/3}}{5^{1/2}D c^4} \left[ 1- \frac{5}{2}e^2 + \frac{35}{24}e^4 + O(e^6)\right]
    \label{eq:h2}
\end{equation}
\cite{Robson2019,Flanagan1998,Finn2000}.
 This expression is obtained using the strain amplitude in the $n$th orbital harmonic $h_n \propto g(n,e)^{1/2}/n$, with $g(n,e)$ given by Equation (20) in \cite{Peters1963}.
For binaries  upto $e \leq 0.2$,  the correction for the eccentricity is less  than  $\sim$10 percent.

In a gravitational wave detector with a sensitivity curve $S_n(f)$, the angle averaged signal-to-noise ratio $\bar{\rho}_2$  over an observing time $T$ is given by
\begin{equation}
    \bar{\rho}_2 = \frac{h_2}{S_n(f)^{1/2}} \sqrt{T}.
    \label{eq:SNR}
\end{equation}
 We put the noise curve of LISA by $S_n(f) = S_d(f)+S_c(f)$ with the detector noise $S_d(f)$ and astrophysical foreground confusion noise $S_c(f)$  \citep{Robson2019}, {where the latter is a function of $T$ \cite{Seto2019}}.

Now we will calculate some estimates related to gravitational wave detection of Galactic DNSBs. Taking a circular binary with the chirp mass $1.2M_\odot$, conservatively located at 20kpc, Equations~(\ref{eq:h2}) and (\ref{eq:SNR}) are used to obtain $\bar{\rho}_2 = 7.8$ and {452} for $f =$ 1.5 mHz and {50mHz} respectively for an observation time $T=$ 4 years. If the observation time is increased to $T=$ 10 years, these increase to $\bar{\rho}_2 =$ 17 and {715}, respectively.

The gravitational wave frequency derivative, $\dot{f}$, characterises small frequency drifts due to GW emission in Equation~(\ref{eq:fdot}). {Even though we used the approximation that $\dot{f} = 0$ in Equations~(\ref{eq:h2}) and (\ref{eq:SNR}), measuring this small quantity is of paramount importance.}  Specifically, $\dot{f}$ can be used in conjunction with Equation~(\ref{eq:h2}) to constrain the chirp mass $\mathcal{M}\propto {\dot f}^{3/5}$ (also {accounting for} the  measured eccentricity discussed shortly).

In terms of the signal--to--noise ratio $\bar{\rho}_2$ and observation time $T$ (longer than 2yr), the frequency derivative can be measured with resolution \citep{Takahashi2002}:
 \begin{equation}
     \Delta \dot{f} \simeq 0.43 \left( \frac{\bar{\rho}_2}{10}\right)^{-1} T^{-2}.
 \end{equation}
 Using this and Equation~(\ref{eq:fdot}), we roughly estimate the fractional frequency resolution ${\dot{f}/}{\Delta \dot{f}} $ for low eccentricity DNSBs located at 20kpc. If we consider binaries with $f=$ 1.5 and 2mHz, then over a 4 year observational period,  we have ${\dot{f}/}{\Delta \dot{f}} =$ 1.0 and 7.4, respectively. These results are summarized in the {upper} part of Table~\ref{tab:NS-examples}. When the observing time is increased to 10 years,  we obtain ${\dot{f}/}{\Delta \dot{f}} =$ 14 and 110.

Therefore, for a 10 year observational period, the majority of low eccentricity DNSBs will have a measurable $\dot{f}$ accurate to within 10\%. Over 4 years, when $f>$2mHz, $\dot{f}$ is accurate within 15\%. {The fractional chirp mass resolution follows}

\begin{equation}
    \frac{\Delta \mathcal{M}}{\mathcal{M}}\simeq  \frac{3}{5}\frac{\Delta \dot{f}}{\dot{f}}.
    \label{eq:freqres}
\end{equation}
It will be possible to select Galactic DNSBs based on the chirp masses (expected to be narrowly distributed around $\sim 1.2M_\odot$), distinct from the much more numerous WDBs in most cases. {This is due to the expected rarity of mHz WDBs with mass $\mathcal{M}>1M_\odot$ (see  e.g.,  Fig. 1 in \citep{Lamberts2019}).
We will later discuss potential issues related to binaries including high mass white dwarfs.}

\subsubsection{Eccentricity}
Known Galactic DNSBs could have residual eccentricities $e \sim 0.1$ by the time they enter the LISA band \citep{Brown2001},
 even despite the tendency to rapidly circularize by Equations~(\ref{eq:fdot}) and (\ref{eq:edot}) as they evolve towards mHz frequencies.

\begin{table}
\caption{Summary of detection properties of the Hulse-Taylor like DNSB.  The binary is conservatively located at 20kpc, and has a chirp mass of $\mathcal{M}=1.2M_\odot$.
}
\label{tab:NS-examples}
\begin{center}
\begin{tabular}{|c|c|}
\hline

 \textbf{Property} & \textbf{Hulse--Taylor like pulsar}  \\ \hline
$\mathcal{M}$ & 1.2 $M_\odot$\\ \hline
${D}$ & 20 kpc\\ \hline
SNR(1.5mHz, 4yr) & 7.8\\ \hline
SNR({50mHz}, 4yr) & {452}  \\ \hline
SNR(1.5mHz, 10yr) & 17\\ \hline
SNR({50mHz}, 10yr) & {715}  \\ \hline
${ \dot{f}}/{\Delta \dot{f}}$(1.5mHz, 4yr) & 1.0 \\ \hline
${\dot{f}}/{\Delta \dot{f}}$(2mHz, 4yr) & 7.4 \\ \hline
${ \dot{f}}/{\Delta \dot{f}}$(1.5mHz, 10yr) & 14 \\ \hline
${\dot{f}}/{\Delta \dot{f}}$(2mHz, 10yr) & 110 \\ \hline \hline
$e$(1.5mHz)=$e_\mathrm{i}$ & 0.057  \\ \hline
$e$({50mHz})=$e_\mathrm{f}$ & {0.0014} \\ \hline
$T_\mathrm{merge, \ i}$ & 486 kyr \\ \hline
$T_\mathrm{merge, \ f}$ & {42} yr \\ \hline
$T_\mathrm{band}$ & {486 kyr} \\ \hline

\end{tabular}
\end{center}
\end{table}
The Hulse--Taylor binary pulsar \citep{Hulse1975} (HT)  has an eccentric ($e=0.6$) 7.8 hour orbit ($f=$ 0.07mHz), and a chirp mass of $1.2M_\odot$. If we consider an HT--like pulsar with today's properties and evolve it from gravitational wave emission through Equations~(\ref{eq:fdot}) and (\ref{eq:edot}), it will have eccentricity $e=0.057$ upon entering the LISA band at 1.5mHz (see {the} lower part of Table~\ref{tab:NS-examples}).

 For such binaries with small eccentricities $e\ll 1$, whose strain in the second orbital harmonic $f_\mathrm{GW}= 2 f_\mathrm{orb}(=f)$ is given by Equation~(\ref{eq:h2}), the sub-leading strains ($\propto e$) will be present in the first and third orbital harmonics (at $f_\mathrm{GW}=1 f_\mathrm{orb}$ and $3 f_\mathrm{orb}$). Using the same technique to obtain Equation~(\ref{eq:h2}), the strain in the third harmonic at $f_\mathrm{GW} = 3f_\mathrm{orb}$ is approximately
\begin{equation}
    h_3 = \frac{9e}{4} h_2.
    \label{eq:h3}
\end{equation}

 Ignoring changes in $S_n(f)$ at $f_\mathrm{GW} = 3f_\mathrm{orb}$ compared to $f_\mathrm{GW} = 2f_\mathrm{orb}$, the signal--to--noise ratio in the third orbital harmonic $\bar{\rho}_3$ is then:
\begin{equation}
    \bar{\rho}_3 \approx\frac{9e}{4} \bar{\rho}_2.
    \label{eq:rho3}
\end{equation}
For $(f,e) = (1.5\mathrm{mHz},0.057)$, the strain in the third orbital harmonic ($h_3$) is 13 percent of the dominant harmonic ($h_2$) via Equation~(\ref{eq:h3}).
If conservatively located at 20kpc, using Equations~(\ref{eq:SNR}) and (\ref{eq:rho3}) we get  $\bar{\rho}_3 = 1$ for a 4yr observation. Therefore, this eccentric binary will not produce a detectable $\bar{\rho}_3 $. At {$50$mHz}, the eccentricity has decreased by {a factor of 40} ($e = 0.0014$), but the signal-to-noise has increased by a factor of 60 (see Table~\ref{tab:NS-examples}). This gives $\bar{\rho}_3 = {1.4}$, which is also not detectable. However, at intermediate frequencies between the {frequency} boundaries, the HT binary's eccentricity is marginally detectable. For example, at $f=15$mHz, $\bar{\rho}_3=5$.

On the $f$-$e$ plane,  we define ``the detectable region'' as the area where we can make a complete survey  for Galactic DNSBs.
So far, we have mainly discussed DNSBs with relatively small residual eccentricities in the LISA band. The detection for these binaries will  be complete at $f\gtrsim 1.5$mHz  for an observational period of $\sim 10$\,yr.  Meanwhile, we are particularly interested in the DNSBs which formed in the LISA band.  If they have large eccentricities, their higher harmonic strains will be essential for detecting them.

However, on the $f$-$e$ plane,  we simply  put  the detectable region as the rectangular area which is bounded by the inequalities $f_\mathrm{i}\equiv 1.5{\rm mHz}\le f \le f_\mathrm{f}\equiv {50}{\rm mHz}$ and $0\le e<1$, ignoring the eccentricity  dependence for the boundary frequencies $f_{\rm i}$ and $f_{\rm f}$.
{Our choice $f_{\rm f}=50$mHz is somewhat arbitrary, and most of our results below are almost independent of it. }

\begin{figure}
  \centering
    \includegraphics[width=0.5\textwidth]{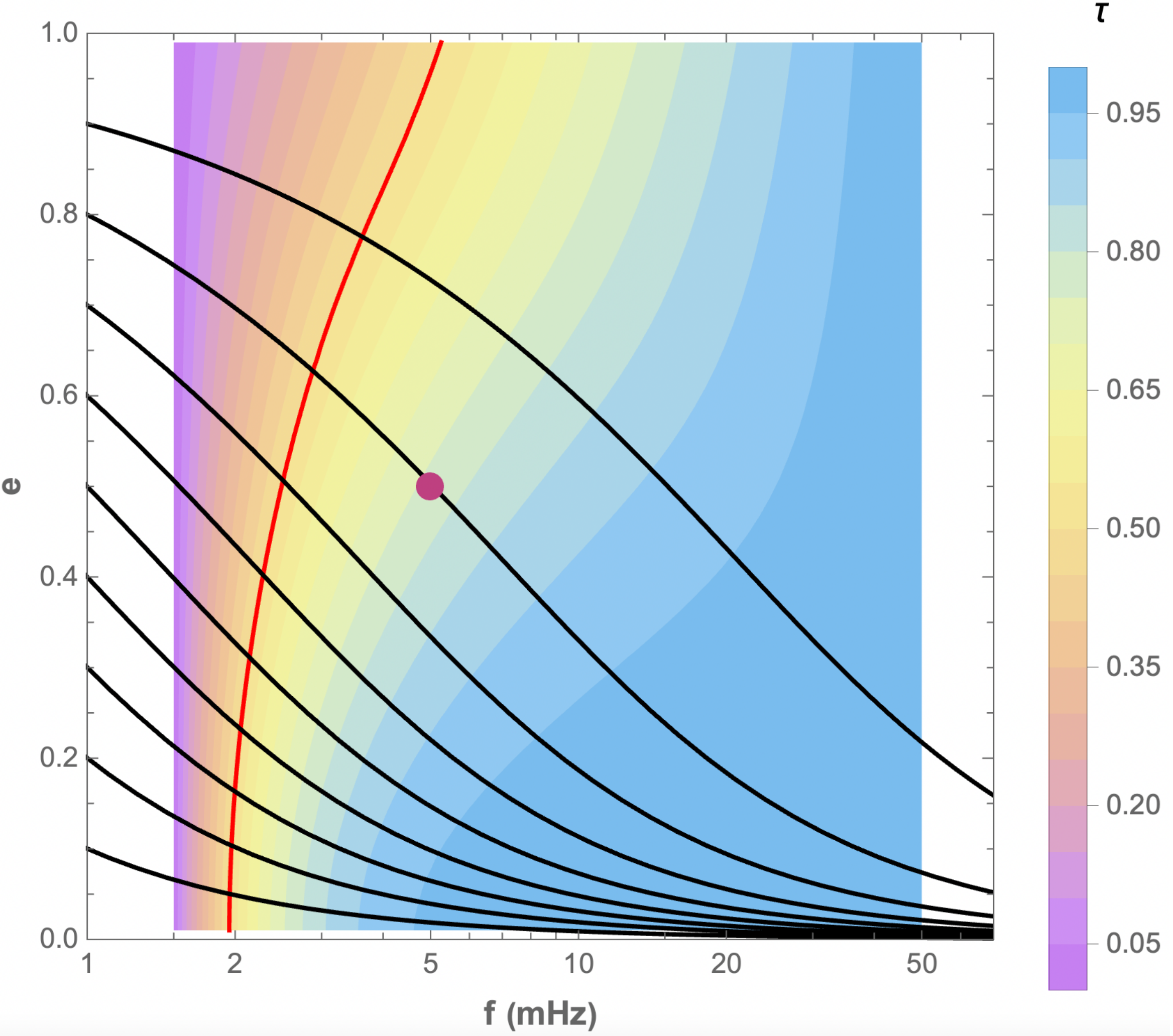}
  \caption{Evolution of Galactic DNSBs on the $f$-$e$ plane.   The detectable region is  shown in pastel colors, bounded by $f_{\rm i}=1.5$mHz and $f_{\rm f}={50}$mHz.  The black curves are the flow lines from Eq. (\ref{eq:dfde}),  moving rightward in time.
  For each point, using the associated flow line, we can define the effective time fraction $\tau(f,e)$, which is shown as a contour map.  We have $\tau=0$ and 1 at the two boundary frequencies.  The red curve corresponds to $\tau = 0.5$ (halfway in this dimensionless time unit). {The purple point on the second flow line corresponds to a binary with $(f,e) =$ (5mHz, 0.5).}}
\label{fig:contour}
\end{figure}

In Fig.~\ref{fig:contour},  the detectable region is shown as the colorful rectangular area.  While it will not be difficult to more precisely include the eccentricity dependence of the boundaries, this task is beyond the scope of our conceptual study. {We hereafter put $N$ as the total number of  DNSBs in the detectable region.}

\section{flow lines in the detectable region}
\label{sec:lisaband}

As discussed  in the previous section, LISA will provide us with Galactic DNSBs scattered in the detectable region on the $f$-$e$ plane. Our primary interest in this paper is to detect the signature of DNSB formation
in the LISA band. {If all of the detected DNSBs were formed  at low frequencies (much lower than mHz) and resultantly  have low eccentricities $e\ll 1$},  the frequency distribution $\mathrm{d} N/\mathrm{d}f$ would be the appropriate data to be analyzed. Indeed, in such a case, we will have the profile $\mathrm{d}N/\mathrm{d}f\propto f^{-11/3}$ as shown in Eq. (\ref{eq:dNdf}).

However, we might detect DNSBs with non-negligible eccentricities. In particular, those formed around the mHz band could have $e=O(1)$. For our study, we thus need to deal with the two dimensional distribution $\mathrm{d}^2N/\mathrm{d}f\mathrm{d}e$.
It is nevertheless advantageous to compress the two dimensional data into much simpler one dimensional data. In fact, there are many useful tools to statistically examine one-dimensional patterns, as we see later.

 The basic question here is whether we have a data compression scheme suitable for studying the potential mHz formation for DNSBs.  In this section, after discussing the long-term evolution of DNSBs in the mHz band, we formally define the effective time fraction $\tau(f,e)$ as a function of $f$ and $e$, for the data compression.

\subsection{Binary evolution in the mHz band}
First, we consider the long-term  evolution of DNSBs in the mHz band due to the GW emission.  From Eqs. (\ref{eq:fdot}) and (\ref{eq:edot}), we have
\begin{equation}
    \frac{\mathrm{d}\ln f}{\mathrm{d}e} = - \frac{18}{19}\frac{(1 + 73/24 e^2 + 37/96 e^4)}{e (1-e^2)(1+121/304e^2)}.
    \label{eq:dfde}
\end{equation}
This equation is independent of the mass parameters.
As mentioned earlier, except for $1-e\ll 1$, the relativistic correction is small for DNSBs  in the LISA band, and  this equation is an  excellent approximation.
We  can easily integrate Eq. (\ref{eq:dfde}) and  obtain the flow lines as shown in Fig.~\ref{fig:contour} with the black lines.
Due to the structure of  Eq. (\ref{eq:dfde}), this family satisfies a self-similar relation.

 For a given point in the detectable region, we can identify an associated flow line. For example, with respect to the  point  $(f,e) =$ (5mHz, 0.5) shown in Fig.~\ref{fig:contour}, we have the flow line that enters the detectable region at $(f_{\rm i},e_{\rm i})=(1.5{\rm mHz},0.74)$  and {escapes it  at $(f_{\rm f},e_{\rm f})=(50{\rm mHz},0.073)$.   Similarly, as presented in Table 1, the HT-like binary would have $e_{\rm i}=0.057$ and   $e_{\rm f}=0.0014$.}

\subsection{Effective time fraction on a flow line}
\label{sec:tau}
 From Eqs. (\ref{eq:fdot}) and (\ref{eq:edot}), we can also estimate the remaining time before the merger \citep{Peters1964}
 \begin{equation}
\begin{split}
    & T_\mathrm{merge}(f,e) = \frac{15}{304}\frac{c^5 }{ \pi^{8/3}G^{5/3}}\mathcal{M}^{-5/3} \\
    & \times f^{-8/3}\left( \frac{1-e^2}{e^{12/19}\left(1+\frac{121}{304}e^2 \right)^{870/2299}} \right)^4\\
    & \times
    \int_{0}^{e} \frac{e'^{29/19}\left(1+\frac{121}{304}e'^2 \right)^{1181/2299}}{\left( 1-e'^2\right)^{3/2}} \ \mathrm{d}e'\,
\end{split}
\label{eq:Tmerge}
\end{equation}
  as a function of $f$ and $e$.
Then, for a given DNSB at $(f,e)$ in the detectable region, we can evaluate the total time $T_\mathrm{total}(f,e)$ that the binary spends on its flow line in the detectable region  ({i.e.} from  the minimum frequency $f_{\rm i}$ to the maximum one $f_{\rm f}$).  More specifically, we can put
\begin{equation}
    T_\mathrm{total}(f,e) = T_\mathrm{merge}(f_{\rm i},e_\mathrm{i}(e,f)) - T_\mathrm{merge}(f_{\rm f},e_\mathrm{f}(e,f)),
    \label{eq:Tband}
\end{equation}
where the eccentricities $e_{\rm i}$ and $e_{\rm f}$ at the two boundary frequencies should be regarded as  functions of $f$ and $e$ (through the corresponding flow line).
{For $f_{\rm f}=50$mHz, the second term in Eq. (\ref{eq:Tband}) is virtually ignorable.}
For the HT-like binary  in Table~\ref{tab:NS-examples},  we obtain $T_\mathrm{total}=$ {486} kyr.
Similarly, the time elapsed after entering into the detectable region at $f=f_{\rm i}$ is given by
\begin{equation}
    T_\mathrm{band}(f,e) = T_\mathrm{merge}(f_{\rm i},e_\mathrm{i}(f,e)) - T_\mathrm{merge}(f,e).
\end{equation}

Then,  we define the effective time fraction
\begin{equation}
    \tau(f,e)   = \frac{T_\mathrm{band}(f,e)}{T_\mathrm{total}(f,e)}
    \label{eq:tau}
\end{equation}
for characterizing  the position of a DNSB on its flow line  in the detectable region.
We note that the effective time fraction $\tau$ is independent of the chirp mass $\mathcal{M}$.

In  Fig.~\ref{fig:contour}, we show the contour levels for the effective time fraction  $\tau$.  We have $\tau=0$ at the lower bound $f=f_{\rm i}$ and $\tau=1$ at the upper end $f=f_{\rm f}$.   The DNSB at  $(f,e) =$ (5mHz, 0.5)  {has $\tau=0.77$}{, shown by the purple point}.
At $e\lesssim 0.05$, the function $\tau(f,e)$ depends very weakly on the eccentricity $e$, as understood from  the weak correction $O(e^2)$ in Eq. (\ref{eq:fdot}).  We can approximately put
 \begin{equation}
{
\tau \approx \frac{f_\mathrm{i}^{-8/3} - f^{-8/3} }{f_\mathrm{i}^{-8/3}-f_\mathrm{f}^{-8/3}} \approx 1-\left( \frac{f}{1.5\mathrm{mHz}} \right)^{-8/3}.}
\end{equation}
at $e\ll 1$.
In the high eccentricity limit $e\to 1$, we have
\begin{equation}
{
    \tau \approx {1.45} \left[ 1 - \left(\frac{f}{1.5\mathrm{mHz}}\right)^{-1/3} \right].}
\end{equation}

For the detectable region in Fig.~\ref{fig:contour}, we have simply ignored the eccentricity dependence of the upper and lower frequency boundaries. Even if the detectable region is deformed on the $f$-$e$ plane, it is straightforward to evaluate the effective time fraction $\tau(f,e)$ by using a family of the flow lines.

\section{statistical tests for in-band binary formation}
\label{sec:injections}
In this section,  we discuss how to examine DNSB injections by using the cumulative distribution function (CDF) of the observed time fractions $\tau$. First,   we derive theoretical expressions for the CDFs without injections and with injections (based on a simple model).
In reality,  due to the finiteness of the sample size $N$,  the observed CDF {will have} fluctuations around the theoretical curves.   We discuss statistical tests to determine the potential injections, under the presence of these fluctuations.

\subsection{Without injections}
Without injections, each binary goes through the detectable region on a flow line at a constant speed $\mathrm{d}\tau/\mathrm{d}t=$ const.  As long as the DNSBs are in a steady state,  after counting contributions from all  flow lines, we have a uniform   probability distribution function (PDF) for $\tau$
\begin{equation}
{\rm Pr}(\tau)={\rm const}.
\end{equation}
 From the normalization condition  $\int_0^1 \mathrm{d}\tau {\rm Pr}(\tau)=1$, we obtain
\begin{equation}
{\rm Pr}(\tau)=1.
\end{equation}
The corresponding  CDF  is given as
\begin{equation}
F(\tau)=\int_0^\tau  \mathrm{d}\tau'{\rm Pr}(\tau')=\tau.\label{eq:CDFuniform}
\end{equation}

\begin{figure}
  \centering
    \includegraphics[width=0.4\textwidth]{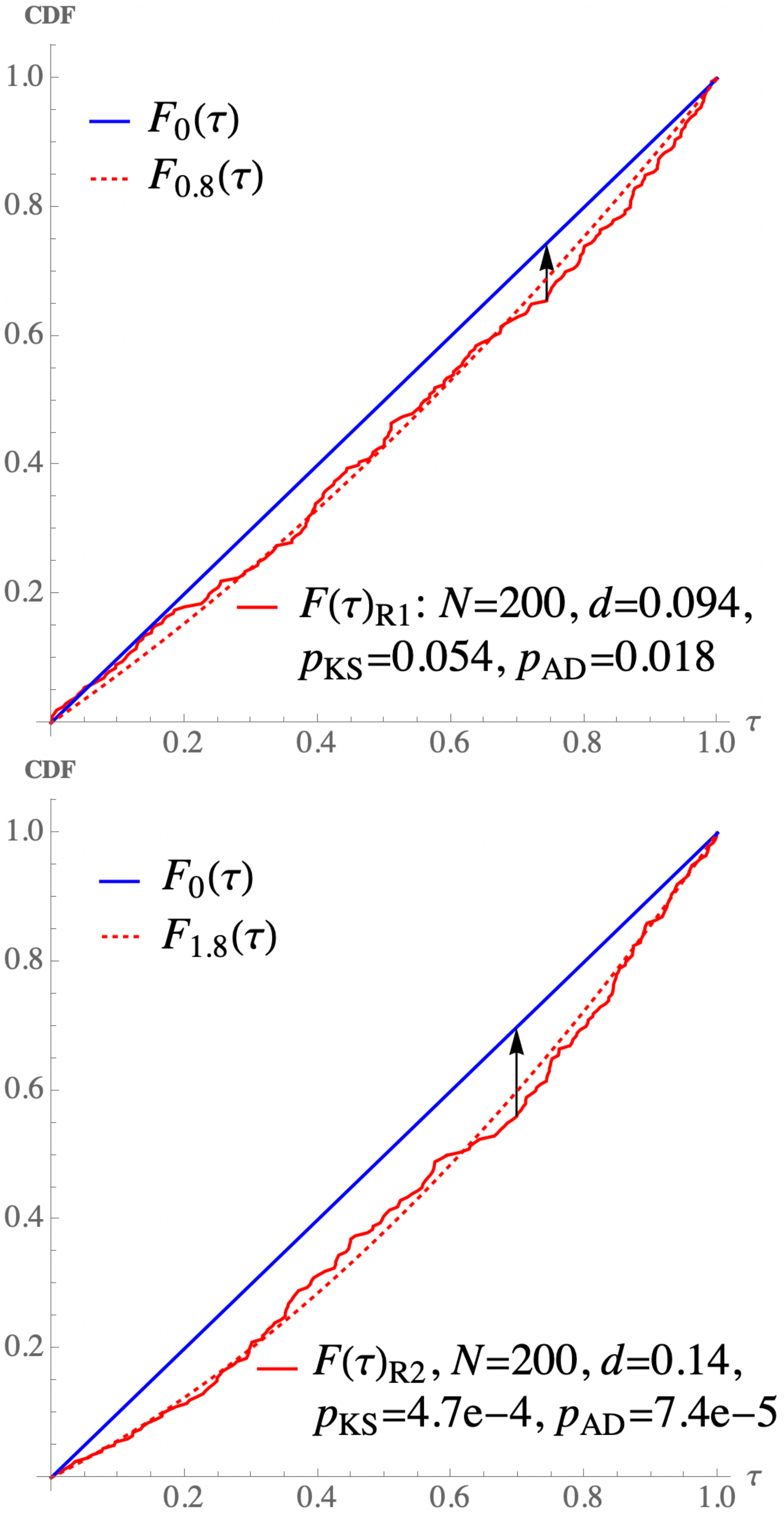}
  \caption{
  The CDFs for the effective time fraction $\tau$, {for the cases of $a=0.8$ and $a=1.8$ are shown in the upper and lower panels}, respectively. The blue curves show $F_0(\tau)=\tau$ without injections (uniform distribution).  The dotted red curves show $F_a(\tau)$ (see Eq. (\ref{eq:cdf})) with the injections parameters $a=0.8$ and 1.8 (or the injected fractions $B=0.29$ and 0.47). The solid  red curves are the corresponding numerical realizations $F(\tau)_{R1}$ and  $F(\tau)_{R2}$  with the sample size $N=200$. The maximum deviations $d$, which are used in  the Kolmogorov--Smirnov (KS) test, are given by the black arrows.  The respective KS $p$-value are 0.054 and $4.7\times 10^{-4}$. Generally speaking, increasing $a$ leads to a larger deviation $d$ and hence smaller $p_{\rm KS}$,  while a larger $N$ gives a curve more similar to the theoretical CDF $F_a(\tau)$.  The $p$-values are also provided for the Anderson-Darling (AD) test. }
\label{fig:single}
\end{figure}

\subsection{With injections}
Now we discuss the PDFs and CDFs { for populations} with injections. As a simple model, we assume a constant injection rate {over $\tau=$ [0,1]} described as
\begin{equation}
\frac{\mathrm{d}{\rm Pr}(\tau)}{\mathrm{d}\tau}={\rm const}.
\end{equation}
After integrating this equation and normalizing the result, we obtain
\begin{equation}
    \mathrm{Pr}(\tau) = \frac{1+a \tau}{1+a/2},
    \label{eq:pdf}
\end{equation}
{where the} positive parameter $a$ is related to the injection rate.
{ In  the numerator of Eq. (\ref{eq:pdf}),  the first and second terms respectively show the component formed below $f_{\rm i}$ and the  component injected in the band $[f_{\rm i},f_{\rm f}]$.}
The associated CDF is given as
\begin{equation}
    F_a(\tau) =  \frac{\tau+a \tau^2/2}{1+a/2},
    \label{eq:cdf}
\end{equation}
and is convex downward.
Here we added the subscript $a$ to explicitly show the parameter dependence.  For $a=0$, we recover Eq. (\ref{eq:CDFuniform}) without injections.

 Note that we do not need to directly deal with detailed models for the eccentricity dependence of  the injections.  This is another advantage of using the compressed variable $\tau$.

{From Eq. (\ref{eq:cdf}) we can easily  confirm that, in the detectable region, the fraction of injected DNSBs in the whole population is given by}
 \begin{equation}
    {B} = \frac{a/2}{1+a/2}.
    \label{eq:bandfrac}
\end{equation}
Since this parameter is more comprehensive than the original one $a$,  we  use them in parallel. Note that for our simple model  (\ref{eq:pdf}), we  have
\begin{equation}
    {B} =\frac12\frac{\mathrm{d}  {\rm Pr}(\tau)}{\mathrm{d}\tau}.
    \label{eq:bandfrac1}
\end{equation}

In Figure~{\ref{fig:single}}, with the red dotted curves,  we show the two  CDFs $F_{0.8}(\tau)$   (upper) and  $F_{1.8}(\tau)$  (lower). We have the injected fractions $B=0.29$ and 0.47 respectively.  We also present the CDF  $F_0(\tau)=\tau$  without injections (blue curves).

\subsection{Statistical tests}
\label{sec:single}
For a {nonzero} injection parameter  $a$, the analytical function $F_a(\tau)$ is distinct from $F_0(\tau)$ without injections. However,  we should recall that the expected number $N$ of the DNSBs in the detectable region is 10's--100's.  If  we make the CDF for the observed  time fractions $\tau $ of this small sample size,  we will have significant scatter due to the finiteness of $N$.

To be concrete, we perform a numerical experiment for $a=0.8$ with the sample number $N=200$.  We employ the scheme known as inverse transform sampling, and generate a list  $\{\tau_{\rm i}\}$ (${\rm i}=1,\cdots,N$) whose individual elements  are drawn from the analytic CDF $F_{0.8}(\tau)$ {without measurement errors}.  After sorting,  we obtain the CDF $F(\tau)_{R1}$, as shown by the solid red line in the upper panel of  Fig. {\ref{fig:single}}. Similarly, we generate another realization $F(\tau)_{R2}$ for $N=200$ and $a=1.8$ (given in the lower panel).   As expected, we can observe fluctuations around the original CDFs $F_{0.8}(\tau)$ and $F_{1.8}(\tau)$ (Eq.~\ref{eq:cdf}) presented with the dotted red curves.

Now, let us assume that LISA provides us with the sample corresponding to  $F(\tau)_{R1}$.  Our central task  here is to statistically determine how likely this  realization  could have been drawn from the model CDF $F_0(\tau)$ without injections (the null hypothesis).
If unlikely, it would be reasonable to claim that we detect a signature of  injections.

There are many sophisticated statistical techniques to check  the consistency of a data sample with respect to a reference CDF. In  the Kolmogorov-Smirnov (KS) test \cite{NumRes}, we evaluate the maximum deviation between  the two CDFs. In our case,  it is expressed as
\begin{equation}
     d =  \max_{0\le \tau\le 1}|F(\tau)_{R1}-F_0(\tau)| .
 \end{equation}

{For an observed sample with size $N$, the probability of obtaining a maximum deviation $\geq d$ is approximately given by \cite{NumRes}}
 \begin{equation}
 \begin{split}
     & p_\mathrm{KS}  \approx \\ &
     2  \sum_{j=1}^{\infty}  (-1)^{j-1}
     \mathrm{Exp}\left\{-2j^2\left[ \left( N_\mathrm{}^{1/2}+0.12+\frac{0.11}{N_\mathrm{}^{1/2}}\right) d \right]^2 \right\}.
     \end{split}
     \label{eq:KSNR}
 \end{equation}
 We use the probability $p_{\rm KS}$ as the $p$-value for the sample $F(\tau)_{R1}$ to be drawn from the CDF $F_0(\tau)$ without injections.

As shown in the upper  panel of  Fig. {\ref{fig:single}}, for the realization $F(\tau)_{R1}$, we have $d=0.094$ and $p_{\rm KS}=0.054$ (using \textsc{Mathematica}).   Similarly, for the other realization $F(\tau)_{R2}$ in the lower panel, we have  $d=0.14$ and $p_{\rm KS}=4.7\times 10^{-4}$.
Note that, in each panel, due to the statistical fluctuations, the black deviation $d$ is larger than the maximum deviation between the two  theoretical curves at $\tau=0.5$. These are respectively given by ${a}/{(4a+8)}=B/4 = 0.071$ and 0.12.

 The KS test is a simple method, and it is most  sensitive to the data around $\tau=0.5$, where we typically have the maximum deviation $d$.  In  contrast, the Anderson-Darling (AD) test uses the whole range of a CDF \cite{NumRes}, and often provides us with a more stringent $p$-value $p_{\rm AD}<p_{\rm KS}$. For the two realizations in  Fig. {\ref{fig:single}},  we  have $p_{\rm AD}= 0.018$ and $7.4\times 10^{-5}$ respectively (again using \textsc{Mathematica}). Interestingly, {for the $a=0.8$ case} the upper sample is now below the standard threshold $p=0.05$.

\subsection{Systematic Study}
\label{sec:results}
We now  systematically explore the parameter space $(N,B)$.
Given the current uncertainties of these parameters,  we consider  the range $0<N<500$ and $0<B<0.5$, as shown in Fig. 3, and divide this space with a 20$\times$20 grid.

\begin{figure}
  \centering
    \includegraphics[width=0.5\textwidth]{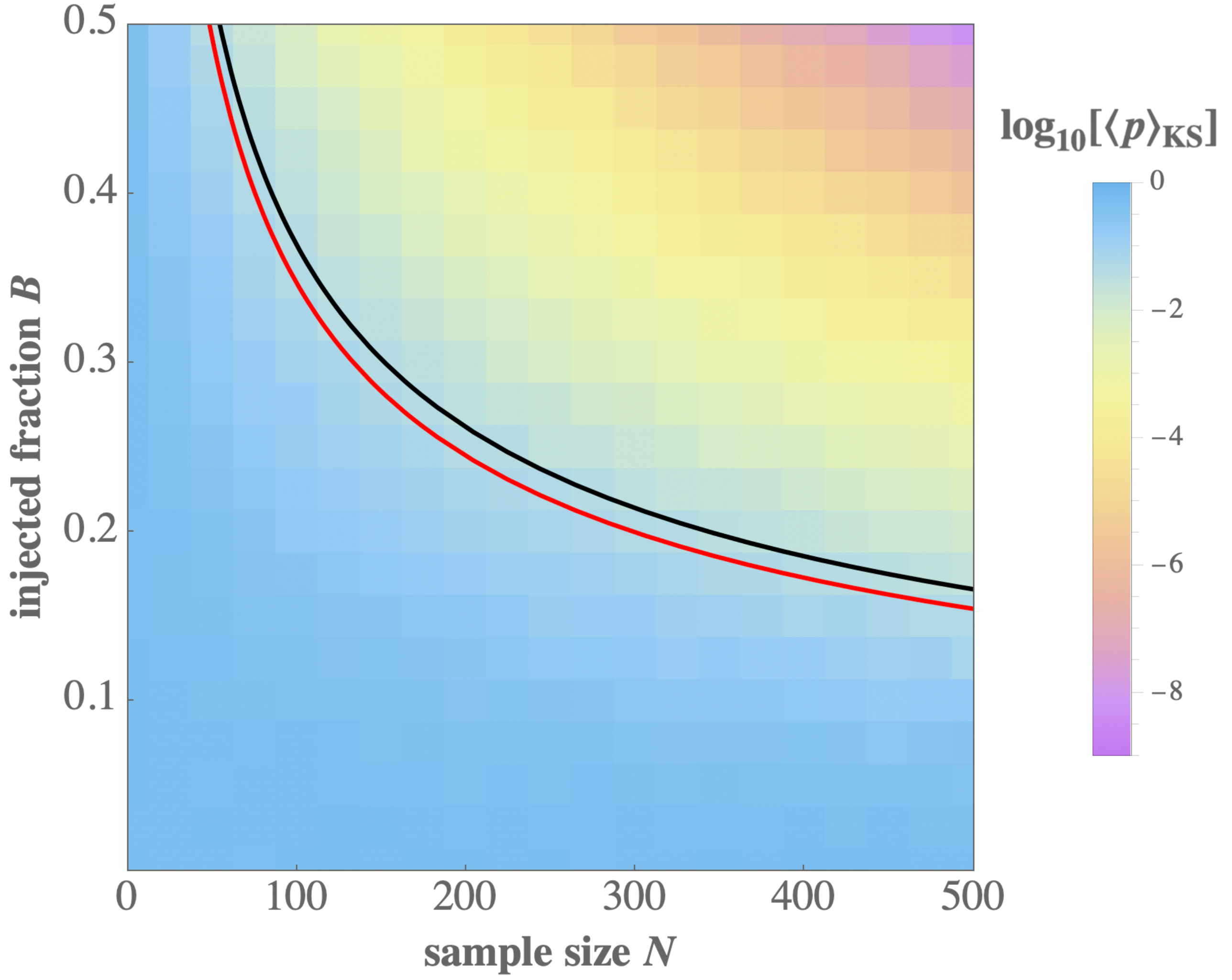}
  \caption{Contour plot of the median $p$-value $\log_{10}[ \langle p
 \rangle_\mathrm{KS}] $ for the Kolmogorov-Smirnov (KS) test. This is calculated for sample size $N$ and injected fraction $B$ with a 20$\times$20 resolution. The black curve shows the  boundary for $\langle p
 \rangle_\mathrm{KS} = 0.05$ (or $\log_{10}[ \langle p
 \rangle_\mathrm{KS}] =-1.3$).  The red curve shows a similar boundary  for the Anderson-Darling (AD) test  $\langle p
 \rangle_\mathrm{AD} = 0.05$. Above these curves, it is likely that the no injection scenario will be rejected due to a $p$-value less than 0.05. }

\label{fig:KS-boundary}
\end{figure}

At each grid point, we simulate 1000  realizations for the $\tau$ distribution,  evaluate the $p$-values  $p_{\rm KS}$ individually for the realizations, and take their median value $\langle p  \rangle_\mathrm{KS}$.
 The numerical results $\langle p
 \rangle_\mathrm{KS}$ are shown in Figure~\ref{fig:KS-boundary} by the pastel levels.

To determine the characteristic curve  corresponding to $\langle p
 \rangle_\mathrm{KS} = 0.05$ in Fig.~\ref{fig:KS-boundary},  we  made iterative calculations  and obtained an approximate expression
\begin{equation}
    B= 0.37 \left(\frac{N}{100}\right)^{-1/2}.
    \label{eq:KS}
\end{equation}
We show this expression with the black curve.
For example, at $B=0.37$, we need $N\sim100$, which  is 3.3 times larger than the reference value {for the expected number of Galactic DNSBs} in  Eq. (\ref{eq:Ng}).

In  the same manner, we  obtain an expression for  the AD test ($\langle p
 \rangle_\mathrm{AD} = 0.05$) as
 \begin{equation}
    B= 0.35 \left(\frac{N}{100}\right)^{-1/2}
    \label{eq:AD}
\end{equation}
presented with the red curve in Fig.~\ref{fig:KS-boundary}.
 For a given injected fraction $B$, the AD test requires $\sim 10$\% smaller sample size $N$ than the KS test.

\subsection{Parameter Estimation}

So far, we have studied the statistical tests to check potential DNSB injections. Here we briefly discuss how  well we can estimate the injected fraction $B$ (or equivalently $a$) from the  CDF of the observed effective time fractions $\tau$.    The standard approach for the estimation is to compare the overall profile of the observed CDF with the theoretical expression (\ref{eq:cdf}) and  fit the parameter $B$ (see e.g., the solid and dotted red curves in Fig.~\ref{fig:single}). Roughly speaking, this method is similar to the concept of the AD test.
Instead,  below, we examine a simple method only using the maximum deviation $d$, as in the case of the KS test.

As mentioned earlier, the maximum deviation $d$ has statistical fluctuations, due to the finiteness of the sample size $N$.  For a given $B$ and $N$, we put the mean value of $d$ by ${\bar d}(B,N)$ and the {root-mean-square} {(rms)} scatter by  $\sigma_d(B,N)$.
If we try  to estimate the parameter $B$ from the observed maximum  deviation $d$ (inversely using the relation $d={\bar d}(B,N)$),  the estimated parameter $B$ contains the rms error   $\sigma_B(B,N)$  roughly given by
 \begin{equation}
    {\sigma_B}(B,N) =  {\sigma_d}(B,N){\left\{  \frac{\partial {\bar d}(B,N)}{\partial  B}\right\}}^{-1}\label{sd}
\end{equation}
with the Jacobian ${\partial {\bar d}(B,N)}/{\partial  B}$.

At various points  in Fig.~\ref{fig:KS-boundary}, we numerically evaluated the two factors $\sigma_d$ and ${\partial {\bar d}(B,N)}/{\partial  B}$ by generating a large number of realizations.  For the scatter $\sigma_d$, we found an approximate relation
 \begin{equation}
\sigma_d(B,N)\simeq0.04 \left(\frac{N}{100}\right)^{-1/2},
\end{equation}
which is independent of $B$.
By taking finite differences instead of the derivatives,  we also found
\begin{equation}
\frac{\partial {\bar d}(B,N)}{\partial  B}\simeq 0.25.
\end{equation}
 This simple relation seems reasonable,  given the maximum deviation $B/4$ between the two theoretical curves in Fig.~\ref{fig:single} (as mentioned earlier).   Then, for the estimation error of $B$, we have
 \begin{equation}
    {\sigma_B}(B,N) \simeq  0.16 \left(\frac{N}{100}\right)^{-1/2}.
\end{equation}
Comparing this result with the critical curves given by Eqs. (\ref{eq:KS}) and (\ref{eq:AD}), we can see that these curves roughly correspond to the condition $B\sim2\sigma_B$.

\section{Discussion}
\label{sec:caveats}
In this paper we have discussed the basic idea {of} probing DNSB injections in the detectable region on the $(f,e)$ plane. Here we mention potential extensions of our study.

\subsection{Other projections}
For simplicity, we proposed the projection  to the single variable $\tau$.  Considering the dimensionality of the original data, we might develop other projection methods, as illustrated in the following example.

Using  the variable $\tau$ in Fig.~\ref{fig:contour}, we first divide the DNSBs into two groups: $G^-$ (with $0\le \tau\le 0.5$)  and $G^+$ (with $0.5< \tau\le 1.0$). Then, along the flow lines,  we move the DNSBs in the group $G^-$,   to the boundary curve $\tau=0.5$ (the red line in Fig.~\ref{fig:contour}).  We put $P^-_{0.5}(e)$ as  the resultant eccentricity distribution    at $\tau=0.5$.   In  the same manner, we can obtain $P_{0.5}^+(e)$ for the group $G^+$.

Now, without the DNSB injections, the two profiles $P_{0.5}^-(e)$ and $P_{0.5}^+(e)$ should be the same, except for some statistical fluctuations. With injections, this is generally not the case.  Therefore, we can apply various techniques, such as the two sample KS-test \cite{NumRes}, for probing potential injections.

\subsection{Lower frequency regime}
We have  dealt with DNSBs only in the detectable region at $f\ge 1.5$mHz, by evaluating the signal-to-noise ratios at a conservative distance of 20kpc.  However, it would be highly desirable to probe the injections in the lower frequency regime.  In the extended regime, LISA's Galactic DNSB survey is  not complete.  For applying our method,   we thus need to correct for the selection bias, by using appropriate models for the  distribution of Galactic DNSBs.

\subsection{Massive white dwarfs }
PSR B2303+46 \citep{vanKerkwijk1999, Tauris2019} and J0453+1559 \cite{Martinez2015} are Galactic neutron star white dwarf (NS-WD) binary candidates with relatively large chirp masses of $\mathcal{M}=1.05 M_\odot$ and $\mathcal{M}=$1.17$M_\odot$ respectively. While they have orbital periods of days (merger timescales larger than Hubble time), we might actually detect similar NS-WD systems with DNSB--like chirp masses in the mHz band. This is because NS-WD are likely to be the next most abundant compact binary after WDBs \citep{Nelemens2001}. {A binary's} white dwarf component could be confirmed  with targeted optical/IR followup, but these searches {may not be successful. Observational challenges include interstellar extinction, and limitations in the sky localisation with LISA.}  In relation to the electromagnetic wave observation, radio followup will be also important for estimating the age of a DNSB \citep{Kyutoku2019}.  Here, we should {note} that a dynamically formed DNSB may have an age largely different from  the spin-down age of the component pulsars.

{In any case, w}e can rather decrease the chirp mass threshold to add  NS-WDs and WDBs  to our binary sample, and thereby increase the total number $N$.  However, in Fig.~\ref{fig:contour},  the flow lines and the associated time fraction $\tau$  are  based on the point particle approximation in Eqs. (\ref{eq:fdot}) and (\ref{eq:edot}).  We thus need to carefully examine the possible modifications induced by the finiteness of  white dwarfs \citep{McNeill2021,Wolz2021}.

\section{Summary}
\label{sec:summary}

Due to the complex interplay between various physical processes, it is still difficult to solidly predict the  orbital parameters  of compact binaries at their formation. New observational results will thus help us to refine theoretical modelings.  In  the near future, the space interferometer LISA will explore GWs around 0.1-100mHz and will make a complete survey for Galactic DNSBs above $\sim 1.5$mHz after $\sim$10yr operation.

In this paper, we discussed how well we can detect  a signature of DNSB formation (injections) around 1mHz with LISA.
We  first introduced the effective time fraction $\tau$ for each DNSB to compress the original two-dimensional data $(f,e)$, as shown in Fig.~\ref{fig:contour}. The probability distribution  function ${\rm Pr(\tau)}$ for the {measured} fractions $\tau$ plays the central role in our method. {There was} a flat profile $\mathrm{d}{\rm Pr}(\tau)/\mathrm{d}\tau=0$ without injections, but $\mathrm{d}{\rm Pr}(\tau)/\mathrm{d}\tau>0$  with injections.

In reality,  we need to  discriminate the differences between the profiles under the existence of {scatter} due to the finite sample size $N$. To be concrete, we made a simple model for the DNSB injections with
$\mathrm{d}{\rm Pr}(\tau)/\mathrm{d}\tau=2B(={\rm const})$, characterized by the injection fraction $B$.  Then, we examined the prospects for discriminating the profile differences with the  Kolmogorov--Smirnov and Anderson--Darling tests.
Our main results are presented in Fig.~\ref{fig:KS-boundary} with the characteristic relations (\ref{eq:KS}) and (\ref{eq:AD})  for the $p$-value of 0.05.  For example, with the sample size $N=100$, we need an injection fraction $B\gtrsim 0.35$ {to detect} injections.

In this paper, we have discussed the very basic idea of studying potential DNSB formation around 1mHz. Our approach can be extended in various directions, including those mentioned in section V.

\section*{Acknowledgements}
LM acknowledges financial support from the Japan Society for the Promotion of Science (JSPS) International Research Fellow program (Graduate School of Science, Kyoto University JSPS P21017). NS is supported by JSPS Kakenhi Grant-in-Aid
for Scientific Research (Nos. 17H06358 and 19K03870). \textsc{Mathematica} was used to perform calculations and generate figures in this work.


\bibliographystyle{apsrev}
\bibliography{apssamp}


\appendix

\label{lastpage}
\end{document}